# Probing Electrocatalytic Gas Evolution Reaction at Pt by Force Noise Measurements. Part 1-Hydrogen

Nataraju Bodappa[1], Zixiao Zhang[1], Ramin Yazdaanpanah[1], Wyatt Behn[1], Kirk H. Bevan[1], Gregory Jerkiewicz[2], and Peter Grutter[1]

[1] Department of Physics, McGill University, 3600 rue University, Montreal, Quebec, Canada H3A 2T8

[2] Department of Chemistry. Queen's University, 90 Bader Lane, Kingston, Ontario, Canada K7L 3N6

**ABSTRACT:** Electrocatalytic processes occurring at the heterogeneous interface are complex and their understanding at the molecular level remains challenging. Atomic force microscope (AFM) can detect force interactions down to the atomic level, but so far it has been mainly used to obtain *in-situ* images of electrocatalysts. Here for the first time, we employ AFM to investigate gas evolution at a platinum ultramicroelectrode (Pt UME) under electrochemical conditions using the force noise method. We detect excess force noise when the individual $H_2$ gas bubble nucleation, growth, and detachment events occur at the Pt UME. The excess noise varies linearly with the applied potential on a semi-log plot. Chronoamperometry current fluctuations indicate that the $H_2$ gas bubble radius increases from $1\pm0.5$ to $10\pm6$ μm and the AFM deflection signal measurements indicate that the mean $H_2$ bubble radius is $321\pm27$ μm. These two values point to the coalescence of small bubbles generated at the Pt UME interface. The contribution reports an innovative method to probe the gas-bubble dynamics at the heterogenous interface under electrochemical conditions with high sensitivity.

## INTRODUCTION

Understanding mechanistic aspects of electrocatalytic reactions at the molecular level contributes to improving the energy efficiency of electrochemical technologies. Important energy conversion and storage technologies, such as water electrolysis produce gas bubbles ($H_2$, $O_2$) on cathode and anode surfaces. Gas bubbles attached to the electrocatalytic surface cover some of its active sites, thus reducing the reaction efficiency and contributing to ohmic losses. Therefore, understanding the nucleation, growth, and detachment of gas bubbles at electrocatalytic interfaces is critical for the continual improvement of electrolyzers' and their energy efficiency.[1, 2]

Gas bubble nucleation and growth at electrocatalytic interfaces have been studied by voltammetry[3], chronoamperometry[4, 5], and optical methods [2,6]. White et al. measured the nucleation rate, the activation energy, and the critical size properties commensurate with single $H_2$ bubble growth.[7] The process was studied using nanoelectrodes 50-100 nm in diameter, wherein it was found that 50 molecules were required to form a hydrogen bubble nucleus[8] with an activation energy of 4-20 $k_B T$.

High-speed shadowgraphy experiments in conjunction with current transient measurement were able to demonstrate that the current oscillations due HER are related to the release of individual $H_2$ bubbles from the surface of a 100 μm diameter Pt microelectrode in acidic electrolyte.[9] Further studies determined that the bubble growth with time (*t*) follows a power law, $R(t) = \beta t^x$, where $R$ is the radius of the bubble, β is the bubble growth coefficient, and *x* is the experimentally determined power value ($x = 0.5$ to 1).[10, 11] Moreover, these experiments revealed that a carpet of ~1-10 micrometer size bubbles exists adjacent to the Pt UME and on top of that another phase of 100-300 μm bubble growth occurs through the coalescence of smaller bubbles from the carpet. [2, 5, 11, 12] The observed current oscillations are related to the larger size bubble detachment.

In this context, it was hypothesized that the bubble growth and detachment stages are governed by a balance between the buoyancy, electrical, and Marangoni forces.[11-13] The bubble detaches from the electrode surface only when the buoyancy force overcomes the electrostatic and Marangoni force.[12] However, direct measurement of these forces during the growth of bubbles is challenging and, thus remains elusive.[14]

It is well established that AFM offers exceptional force sensitivity, down to atomic-level interactions, and is capable of imaging surfaces with atomic resolution even under electrochemical conditions.[15] However, experimental work employing AFM to study bubbles at electrocatalytic interfaces is challenging, and, consequently, only a few reports exist. [16] For example, *in-situ* AFM was successfully used to map static bubbles.[17] Yet, bubble dynamics, which correlate the force and force noise to the bubble growth and detachment, were not investigated by AFM. Results of such experimental work are urgently required and would lay the foundation for testing theories of bubble growth and detachment forces, as well as identifying critical parameters to advance the understanding of bubble dynamics and behavior at electrocatalytic interfaces. [11-13,18,19]

In this contribution, we applied for the first time *in-situ* AFM to analyze the time-dependent bubble dynamics at the surface of Pt microelectrodes during the $H_2$ evolution reaction (HER) at different overpotentials. We utilize high-speed digitalization of AFM measurements to acquire force noise power spectral density(PSD) properties concurrently with electrochemical measurements. Analysis of the results sheds important light on $H_2$ bubble dynamics (growth and detachment) in relation to the overpotential. Examination of individual current and force fluctuations in the current and force transients facilitates the determination of the surface charge and buoyancy force associated with the formation of an individual $H_2$ bubble, thus its size and the number of $H_2$ molecules therein. Our results are in good agreement with those reported using high-speed shadowgraphy,[2, 11] thus representing a novel complementary approach towards the analysis of $H_2$ bubble dynamics at Pt electrodes (and many other related systems).

## RESULTS AND DISCUSSION

Figure 1a presents schematics of the experimental set-up used to acquire force noise power spectral density (PSD) while conducting electrochemical measurements. To measure force noise associated with the bubble nucleation, growth, and detachment during the HER, a Pt-coated AFM cantilever (details in SI) was used. Electrochemical measurements were performed using a three-electrode configuration consisting of a Pt ultramicroelectrode (UME) as a working electrode (WE), a loop-shaped Pt counter electrode (CE), and a Pt wire that served as a quasi-reference electrode using an external potentiostat (Figure 1a; details in SI). Conversion of the potential values from the Pt quasi-reference electrode to a reversible hydrogen electrode (RHE) is discussed in the SI. The potential of the quasi-reference electrode was determined to be –1.1 V on the RHE scale in the same electrolyte solution (Figure S1). The potential of the WE was controlled using an external potentiostat. All potential values are reported on the RHE scale. Figure 1b shows a cyclic voltammetry profile of the disk-shaped Pt UME (WE) in 0.10 M aqueous $K_2SO_4$ solution in the potential range of the HER and double-layer charging. As expected, the magnitude of the negative current ($I$) in the negative-going transient increases exponentially due to the HER. In the positive-going transient, the peak at ca. 0.1 V is due to the $H_2$ oxidation reaction (HOR) and the features at higher potentials are due to the oxygen reduction reaction (ORR) and surface oxide formation.[20] It is important to emphasize that the measurements were conducted under ambient conditions (at $T$ = 293 K and exposure to air), hence in the double-layer potential region there is a contribution associated with the ORR. Once the HER commences, the contribution of the ORR to the overall current is negligible because the exchange current density ($j_o$) of the ORR is orders of magnitude lower than that of the HER.[21]

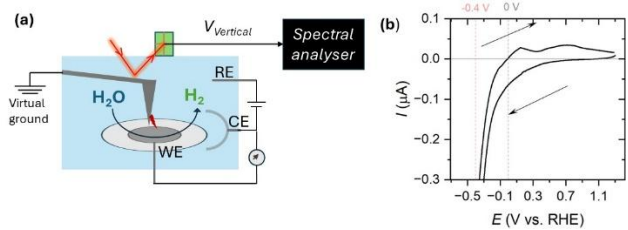

**Figure 1.** Schematic representation of electrochemical (EC)-AFM setup used to measure the force noise due to hydrogen evolution reaction (a); an external potentiostat is used to drive catalytic reactions at the Pt microelectrode. (b) CV of Pt UME in 0.1 M $K_2SO_4$.

Figure 2a represents chronoamperometry (CA) transients for the WE when its potential is stepped from the open circuit potential (1.1 V vs. RHE, Figure S1) to a potential in the 0.3 V to –0.6 V range. In all transients, we observe an initial drop in the value of the current ($I$) that then levels off and reaches a steady-state current ($I_{ss}$). When the potential is stepped from an initial value ($E_i$) to a final potential ($E_f$) that is higher than 0.0 V, then $I_{ss}$ is in the nA range. However, when $E_f$ is lower than 0.0 V, then $I_{ss}$ is significantly higher (ca. –0.1 µA for $E_f$ = –0.30 V and ca. –2 µA for $E_f$ = –0.6 V). We emphasize that when –0.6 ≤ $E_f$ ≤ –0.4 V, then noise in the form of spikes ('blips') is observed on the $I_{ss}$ vs. time ($t$) transients (the inset, Figure 2a). The sharp drop within an individual current blip (to a more negative value) is attributed to the detachment of $H_2$ gas bubbles. On the other hand, the current increase (to a less negative value) within an individual blip is ascribed to the nucleation and growth of an $H_2$ gas bubble, similar to the observations reported in an acidic solution.[2, 11]

Henry's law relates the amount of dissolved $H_2$ gas (expressed as a mole fraction, $x_B$, or a molar concentration $c_B$) to its partial pressure ($p_B$) above the liquid, $p_B = x_B K_{H,x}(B)$ and $p_B = c_B K_{H,c}(B)$, where $K_{H,x}(B) = 7.1 \times 10^4$ atm and $K_{H,c}(B) = 1.3 \times 10^3$ atm mol$^{-1}$ L are the respective Henry's law constants for $H_2$ gas.[22] Even in the case of $p_B = 1.0$ atm, the concentration of dissolved $H_2$ gas is only $c_B = 7.7 \times 10^{-4}$ mol L$^{-1}$, thus very low. Although we conduct experiments using an open cell, a supersaturation state is established when $E_f$ overcomes a threshold limit, which in our case is –0.4 V ≤ $E_f$ ≤ –0.3 V (Figure 2a).[2]

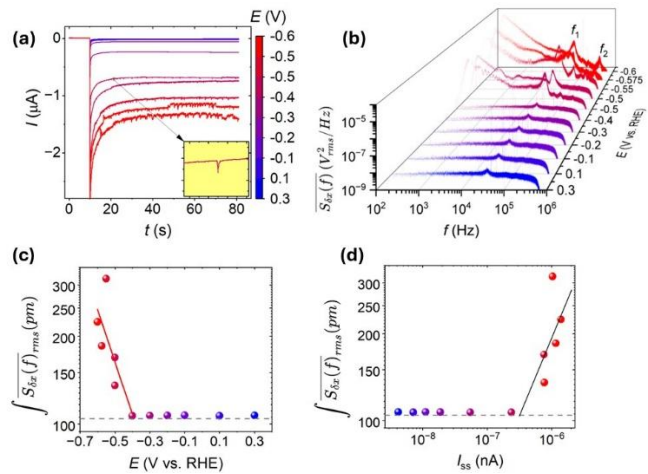

**Figure 2.** (a) Current transient response of Pt UME for different final potential ($E_f$) steps from open circuit potential. Inset shows the "blip" response for hydrogen bubble detachments. (b) Average deflection noise power spectral density, $\overline{S_{\delta x}(f)}$ of AFM cantilever measured in response to corresponding $E_f$ shown in panel (a). The bias voltage coloring is the same in all panels. Tip-sample distance was set to 15 nm. (c) Correlation of integrated deflection noise versus electrochemical potential on a log-linear plot. (d) Correlation of integrated deflection noise versus electrochemical steady-state current on a log-log plot. Sampling rates: 2.5 MHz, measurement bandwidth 1 MHz.

Simultaneous to the CA measurements, the cantilever's average deflection noise PSD ($\overline{S_{\delta x}(f)}$) was collected for each $E_f$ value (Figure 2b). Notably, in the case of $E_f$ ≤ –0.4 V $\overline{S_{\delta x}(f)}$ shows a substantial increase in the noise level (Figure 2c), which we refer to as "excess noise" because it significantly exceeds the average noise level intrinsic to the experimental set-up. We define the "excess noise" as the frequency-integrated average deflection noise ($\int \overline{S_{\delta x}(f)}$) above the cantilever background at open circuit potentials (PSD integrated from 250 Hz to 0.6 MHz). The background noise is dominated by the well-understood AFM detection noise and thermal cantilever noise,[23] visible as the thermally driven cantilever resonance peak at $f_1 = 41.641$ kHz.

Figure 2c shows the relation between the frequency-integrated noise $\overline{\int S_{\delta x}(f)}$ and the applied $E_f$. In the case of $-0.4$ V $\leq E_f \leq 0.3$ V, the value of $\int \overline{S_{\delta x}(f)}$ remains constant (ca. 105 pm) and is limited by the instrumental noise floor measured at the open circuit potential. In the case of $E_f \leq -0.4$ V, the value of $\int \overline{S_{\delta x}(f)}$ increases linearly with decreasing $E_f$ in a semi-logarithmic plot with a slope of $-1.8\pm0.7$, thus indicating that the average bubble size increases. Figures 2a and 2b point to $H_2$ gas bubble formation already at $E_f = -0.4$ V, which, however, are not detected through the analysis of the $\int \overline{S_{\delta x}(f)}$ vs. $E_f$ plot (Figure 2c). There are two possible explanations for this behavior: (i) bubble generation is not recorded because we use a high-pass filter (200 Hz), and the bubble generation rate is lower than this frequency; and (ii) bubbles formed are very small and the force signal associated with their detachment is smaller, which does not generate enough signal strength to record a noise PSD.

Figure 2d shows a plot of the $\int \overline{S_{\delta x}(f)}$ vs. $I_{ss}$ on a log-log scale, which reveals two trends. In the case of an absolute value of $I_{ss}$ being up to 55 nA ($|I_{ss}| \leq 55$ nA), thus for $E_f \geq -0.4$ V, the value of $\int \overline{S_{\delta x}(f)}$ remains constant (ca. 105 pm). However, in the case of $|I_{ss}| > 55$ nA), thus for $E_f \leq -0.5$ V, the value of $\int \overline{S_{\delta x}(f)}$ increases linearly on the log-log scale with a slope of $0.5\pm0.3$.

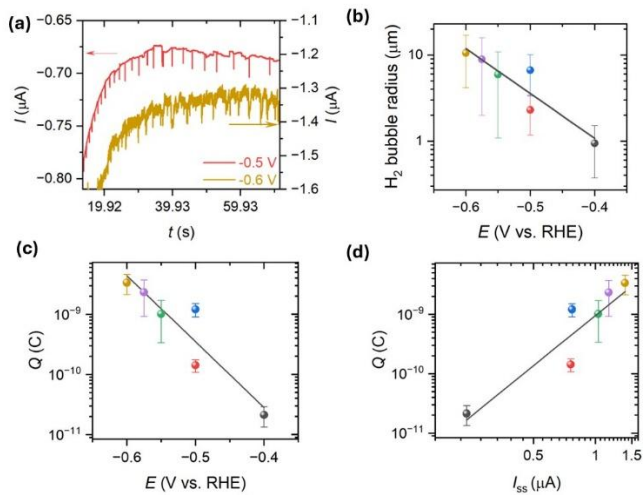

**Figure 3.** (a) Current-time trace of Pt UME subjected $E_f = -0.5$ V, and $-0.6$ V vs. RHE from open circuit potential. (b) $H_2$ gas bubble radius vs. $E_f$. The average blip charge involved the nucleation, growth, and detachment of the $H_2$ bubble as a function of (b) potential and (c) steady-state current for HER on a log-linear plot and log-log plots. (d) Average deflection noise PSD of AFM cantilever at $-0.6$ V and $-0.7$ V within 200-4 kHz range.

The HER commences as soon as the applied $E_f$ is negative on the RHE scale but the blips in the $I$ vs. $t$ transients regularly appeared only in the case of $-0.6$ V $\leq E_f \leq -0.4$ V (Figure 3a). As explained above, they are due to the individual $H_2$ gas bubbles generated on the surface of Pt UME when the amount of the $H_2$ gas being generated exceeds the solubility level determined by Henry's law (a supersaturated solution develops).

Each blip-shaped response is attributed to the nucleation, growth, and detachment of an $H_2$ gas bubble. An $H_2$ gas bubble being formed blocks a part of the electrode surface and reduces the current. Its detachment does the opposite, thus unblocking the electrode surface and increasing the current.[11] Integration of the "missing current" in the $I_{ss}$ vs. $t$ transient for an individual blip allows us to determine the charge ($Q$) associated with a temporary removal of a section of the double layer when an $H_2$ gas bubble is present growth occurs. Assuming the spherical shape of an individual $H_2$ gas bubble, the area where a double layer is removed has a circular shape. The surface area of a 2D projection of a spherical $H_2$(g) bubble requires knowledge of the double-layer charge density ($q_{DL}$) and the value of $Q$. The approach adopted to determine $q_{DL}$ and $Q$ values is visualized in Figures S2.

Figure S2a presents a typical $I$ versus $t$ transient when the applied potential is stepped from the open circuit potential (1.1 V vs. RHE) to a final potential in the $-0.60 \leq E_f \leq 0.30$ V ($E_f = -0.10$ V in Figure S2a). The $I$ vs. $t$ transient reveals a sharp drop over the initial 10 ms that is due to the double layer charging; the steady-state current that eventually establishes is attributed to the ORR (in the case of $0.00$ V$\leq E_f \leq 0.30$ V) or the HER (in the case of $E_f \leq -0.10$ V). Integration of the $I$ versus $t$ transient over the initial 10 ms yields the charge of the double layer charging ($Q_{DL}$). Knowing the diameter of the Pt UME ($d = 25$ μm) and its surface roughness ($R = 2$), we determine the electrochemically active surface area ($A_{ecsa}$) of the electrode; subsequently, we determine the double layer charge density, $q_{DL} = Q_{DL} / A_{ecsa}$. These calculations are performed only for the $0.00$ V $\leq E_f \leq 0.30$ V range because we do not want to have a contribution from the HER to the $Q_{DL}$ determination, and then we make a plot of $q_{DL}$ vs. $E_f$ (Figure S2b). The relationship is linear and its extrapolation to the potential at which blips are first observed ($E_f \leq -0.40$ V) yields the double-layer charge density in the presence of $H_2$ gas bubbles. Knowledge of the "missing charge" $Q$ and the double layer charge density $q_{DL}$ determined using the above-described approaches enabled calculation of the $H_2$ gas bubble radius $r$ for each value of $E_f$ using $r = \sqrt{\left(Q/\pi q_{DL}\right)}$ (its derivation is shown in SI). The calculations show that the $H_2$ bubble radius is in the 1–10 μm range and increases as $E_f$ increases from $-0.4$ V to $-0.6$ V (Figure 3b, details in SI, Table S1 and S2).

A comparison of the chronoamperometry transients (Figure 3a presents two examples; Figure S3 presents results for six $E_f$ values) reveals the periodicity of the blips associated with the detachment of $H_2$ gas bubbles. The results demonstrate that as $E_f$ becomes more negative, the number of blips increases. The results allow us to determine the number of detachments of $H_2$ gas bubbles per unit of time, thus the event's frequency; it is found to be in the 1–10 Hz range (Figure S4). Since the scale in Figure S4 is semi-logarithmic, the rate of $H_2$ bubble detachments increases exponentially with $E_f$, thus following a Butler-Volmer like kinetics relationship. Our results obtained for a Pt UME in neutral media are very similar to those acquired for a Pt UME in an acidic electrolyte solution.[2]

A semi-logarithmic plot of the average bubble blip charge, $Q$ (determined by integrating the "missing current" in the $I_{ss}$ vs. $t$ transient for an individual blip) vs. $E_f$ is linear and has a slope of $-10.9\pm3.6$ V$^{-1}$ further revealing that the hydrogen bubble size increases exponentially (Figure 3c). The results presented in Figures 2a, 3a, 3b, and S3 lead to the observation that $Q$ increases with $I_{ss}$ yielding a power law relationship with a slope of 2.8 ± 0.9 (Figure 3d). The trends reported in Figures 3c and 3d also resemble the trends presented in Figures 2c and 2d. Therefore, our results further support the proposal that the excess force noise originates from the $H_2$ gas bubbles' interaction with the AFM cantilever.

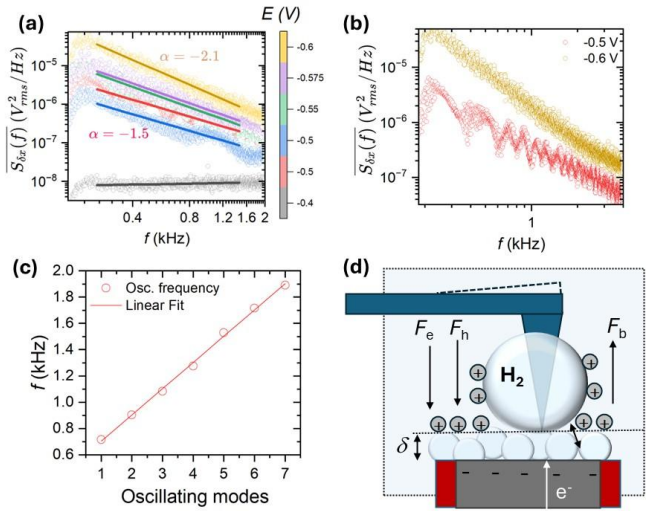

**Figure 4.** Average deflection noise PSD of AFM cantilever for (a) $E_f$ = –0.4, to –0.6 V within 200-2 kHz. (b) $E_f$ = –0.5 and –0.6 V within 200-4 kHz (c) Oscillation peak frequencies extracted from panel b, for $E_f$ = –0.5 V. (d) Schematic of an individual $H_2$ bubble detachment and different forces exerted at gas evolving Pt UME.

Figure 4a represents the plot of low-frequency force noise PSD $\overline{S_{\delta x}(f)}$, for < 2.0 kHz) for –0.40 V ≤ $E_f$ ≤ –0.60 V. The plots for −0.60 ≤ $E_f$ ≤ –0.50 V follow a linear dependence on a log-log scale with a power law relationship, $\overline{S_{\delta x}(f)} = Cf^\alpha$. The value of $\alpha$ transitions from –1.4 for $E_f$ = –0.50 V to –2.0 for $E_f$ = -0.60 V. The plot for $E_f$ = –0.40 V stands out in the sense that it is almost parallel to the frequency axis. This behavior is consistent with our analysis of the results presented in Figures 2c, which imply that there is no noise because there are practically no $H_2$ gas bubbles formed.

Figures 4b, 4c, and S5 show low-frequency noise PSD plots for -0.50 V and -0.55 V reveal force noise discrete oscillations with an average frequency of 200 Hz (~ 5 ms) (Figure 4b and S5). These oscillations are due to the competition between the buoyancy force ($F_b$) and the electrostatic force ($F_e$). (Figure 4d) In the case of less-negative $E_f$ values, the electrostatic force predominates because the $H_2$ gas bubbles are small, and in the case of more-negative $E_f$ values, the buoyancy force predominates because the $H_2$ gas bubbles are large. A similar $H_2$ gas bubble oscillations were computed from the data obtained using high-speed shadowgraphy experiments.[11]

Elsewhere, it was reported that the coalescence of small droplets into larger ones gave rise to a power law relationship with $\alpha = -2$.[24] As in our case the bubble growth also follows a power law relation with $\alpha = -2$ at the most negative potential ($E_f$ = –0.60 V). The similarity between the theoretical calculations and our experimental results points to the possible coalescence of small $H_2$ gas bubbles to larger ones as shown in schematics (Figure 4d) This proposal is consistent with high-speed shadowgraphy measurements in aqueous acidic electrolytes for Pt UME.[2, 11]

From the low-frequency noise force plots we determine the value of $\alpha$ and plot it as a function of the average $H_2$ gas bubble size.(Figure S6) The absolute values of power law exponent, $\alpha$ and total force noise ($\int \overline{S_{\delta x}(f)}$) increases non monotonically with the average $H_2$ gas bubble radius estimated from the chronoamperometry fits. (Figure S6a and S6b)

Above, it is explained that the background noise is dominated by the AFM detection noise and thermal cantilever noise, visible as the thermally driven cantilever resonance peak at $f_1$ = 41.641 kHz; its second eigenmode is $f_2 = \sim 6 f_1 =$ 249.85 kHz. The ratio of the two fundamental resonance frequencies ($f_2/f_1$) of the AFM cantilever plotted *vs.* $H_2$ bubble radius yields an exponential relationship (Figure S6c). This finding implies that the damping induced by the electrolyte on the cantilever tip becomes reduced by the growth of an $H_2$ bubble beneath the entire cantilever.

Noise measurements with Si cantilevers did not produce the same results as those obtained with Pt cantilevers. In the case of the Si cantilevers, the $H_2$ gas bubble continuously grew and interfered with the AFM optical detection scheme (Figure S7a). Such force noise spectra typically produce broad oscillation peaks below the cantilever's fundamental resonance. Similar phenomena were ascribed to the bubble's thermal fluctuations.[25] However, experiments conducted using low spring constant cantilever ($k$ = 0.3 N/m) (Figure S7b) and tipless Si cantilevers show an excess force noise as a function of potential as early as at $E_f$ = -0.2 V and show $1/f^2$ behavior. (Figure S8).

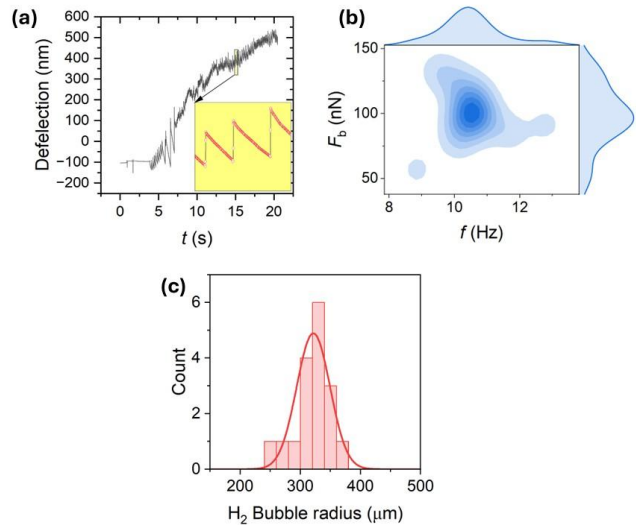

**Figure 5.** (a) AFM cantilever tip-sample interactions recorded as deflection signal at $E_f$ = -0.325 V, Spring constant $k$ =1.33 N/m (b) Total force extracted from the sharp rise in the discrete deflection fluctuations vs. observed frequency. (c) Histogram of the $H_2$ gas bubble radius estimated from the discrete buoyancy forces from panel b.

To investigate the origin of the excess force noise PSD, we also examined the AFM deflection signal tip-sample interactions as a function of time. We noticed that the signal fluctuates during the $H_2$ gas bubble evolution. Figure 5a represents the deflection signal of the Pt cantilever at $E_f$ = -0.325 V. Major forces involved during the growth of $H_2$ gas bubbles and their detachments are the electrostatic ($F_e$), hydrodynamic ($F_h$), and buoyancy ($F_b$), forces.[11, 26] The former two forces act towards the surface of the Pt UME and only the $F_b$ acts in the opposite direction. Therefore, the sharp rise in the deflection signal corresponds to the buoyancy force that originates from the $H_2$ gas bubble detachment (Figure 5a). A plot of the buoyancy forces vs. the frequency of the events yields that $F_b$ = 100 nN with ~10 Hz frequency.(Figure 5b) The slow decay in force signal corresponds to the bubble nucleation and growth. The bubble buoyancy force is $F_b = \frac{4}{3}\pi r^3 \nabla \rho g$; where $r$ is the bubble radius, $\nabla\rho = \rho_l - \rho_g$ refers to the density difference between a liquid electrolyte solution and the $H_2$, and $g$ is the gravitational acceleration.[26] The deflection signal (Figure 5a) can be converted to the deflection force ($F_d$). Because the deflection force equals the buoyancy force ($F_d = F_b$), we can apply the above equation to determine the radius of the $H_2$ gas bubble. Because the deflection adopts a set of values, it leads to a histogram of values of the radius (Figure 5b) and a mean radius of 321.0±27.7 $\mu m$. This value is closely consistent with the bubble radius extracted from the high-speed shadowgraphy experiments in aqueous acidic electrolytes.[2, 5, 11] Note that slight inconsistency in the onset potential for the excess noise observation either comes from the different electrolyte volumes used with an open atmosphere and/or surface reconstruction of the Pt UME under $H_2$ gas evolution.[27]

The mean radius of the $H_2$ gas bubbles determined based on electrochemical measurements is in the 1–10 μm range and increases with $E_f$ becoming more negative. The significantly larger value of the $H_2$ gas bubble determined from the force measurements suggests that smaller bubbles are initially formed and that they quickly undergo coalescence before being measured by AFM, as visualized in Figure 4d.

## CONCLUSIONS

In conclusion, we employed AFM to investigate the force noise of a Pt-coated AFM tip during the $H_2$ gas evolution at a Pt UME and detected force fluctuations as $H_2$ gas bubbles developed and detached at the electrode surface. Excess force noise was observed at the onset of current fluctuations caused by a single $H_2$ gas bubble growth and detachments. The excess force noise shows a linear dependence with the applied potential on a semi-log plot and a linear dependence with the steady-state current on a log-log plot. The $H_2$ gas bubble radius extracted from the chronoamperometry measurements shows that the bubble's average radius size increases from 1±0.5 to 10±6 μm with the applied potential. Detailed investigation of tip-sample force interactions enabled direct characterization of the growth and detachment of individual $H_2$ gas bubbles. The $H_2$ gas bubble radius was measured by equating the sharp rise portion of the measured force fluctuation to the buoyancy force. Force measurements yielded an average size of the $H_2$ gas bubbles of 321.0±27.7 $\mu m$, which is significantly larger than the value obtained from the chronoamperometry measurements. These two radius values indicate that the excess force noise detected by the AFM tip vastly originates from the coalescence of the individual smaller bubbles.

As visualization and quantification of gas bubbles at nanometer dimensions on catalytic surfaces remain challenging, further work using our force noise-based imaging will enable new insights into the bubble formation mechanism. By imaging bubble-forming nucleation sites at the electrochemical interfaces our method could be extended to probe complex nanoscale dynamics of reaction processes at electrocatalytic interfaces. Instrumental detection limits also limit the detection of excess force noise. Further reduction of instrumental noise floor or the reduction of distance separation between tip and sample could enable to study of adsorption and active sites at the reactive interfaces.

## ASSOCIATED CONTENT

### Supporting Information

The Supporting Information is available free of charge on the ACS Publications website.
Details on the experimental methods, data acquisition, and analysis used to present the main text data are provided. Additional supporting data (charge analysis, Z feedback effects to current and deflection noise) is included ( PDF)

## AUTHOR INFORMATION

### Corresponding Author

*__Nataraju Bodappa__ - *Department of Physics, McGill University, Montreal, Quebec, Canada H3A 2T8.*
orcid.org/0000-0003-4387-8887.
*Email: Nataraju.bodappa@mcgill.ca.*

### Authors

**Zixio Zhang** - Department of Electrical and Computer Engineering, University of California, Davis, CA 95616, USA

**Ramin Yazdaanpanah** - Department of Electrical and Computer Engineering, Northwestern University, Evanston, IL 60208, USA

**Wyatt Behn** - Department of Physics, McGill University, Montreal, Quebec, Canada H3A 2T8.

**Kirk H. Bevan** - Department of Mining and Materials Engineering, McGill University, Montreal, Quebec, Canada H3A 2T8

**Gregory Jerkiewicz** Department of Chemistry. Queen's University, 90 Bader Lane, Kingston, Ontario, Canada K7L 3N6

**Peter Grutter** - Department of Physics, McGill University, Montreal, Quebec, Canada H3A 2T8.

### Author Contributions

The manuscript was written with contributions from all authors. All authors have approved the final version of the manuscript.

### Notes

The authors declare no competing financial interest.

## ACKNOWLEDGMENT

This research was supported by. FRQNT Team grant (FRQ-NT PR-298915), NSERC Discovery Grant and NRC COLLABORATIVE RESEARCH AND DEVELOPMENT (RGPIN-2021-02666 and NRC: CSTIP Grant AI4D-131-1)

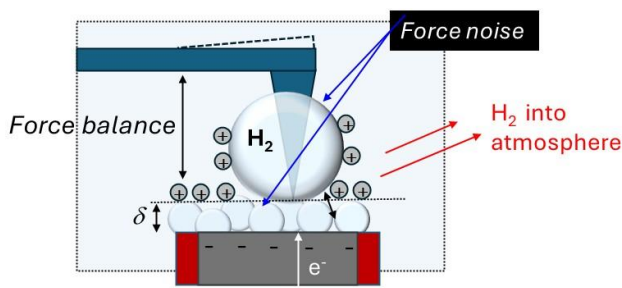

# Supporting Information

**Materials and methods**

An Asylum Molecular Force Probe (MFP-3D) atomic force microscope on an inverted optical microscope with a standard optimized Resistance Conductance Amplifier (ORCA) holder was used to acquire all the data presented in this manuscript. A moderate spring constant (2-6 N/m) cantilever was chosen to achieve the detection limits of the instrument. We used a conductive Pt cantilever (AC240PP; l×b = 240×40 μm purchased from NanoAndMore) with a tip height of 14 μm. A 25-micrometer diameter Pt disk ultra-microelectrode with glass sheath was used as a working electrode. A Pt wire was used as a counter and quasi-reference electrode. $K_2SO_4$ (purity > 97%, Aldrich) was used as received. Milli Q water with an electrical resistivity of 18.5 MΩ. cm was used for all the experiments. The experiments were conducted in an open atmosphere without purging inert gas into the supporting electrolyte.

**Data acquisition and analysis**

Standard calibration was first performed in the air to obtain the spring constant ($k$) and optical lever sensitivity (OLS) of the cantilever. Second, the optical lever sensitivity is further calibrated by using the thermal spectrum obtained in the aqueous electrolyte. A typical deflection inverse optical lever sensitivity of 151.02 nm/V was obtained. Deflection and lateral AFM signals were always kept close to the center of the photodiode to allow for maximal electronic amplification of the noise without saturation of the data acquisition (DAQ) system. If not specifically stated, deflection represents the vertical signal of the quadrant photodiode.

To control the electrochemical potential and measure the cyclic voltammetry, an amperometric response from the Pt UME a CHI1030 potentiostat was used. Pt ultramicroelectrode(UME) was chosen as its size limits the amount of hydrogen generation which inherently limits the hydrogen concentration to grow large-size bubbles rapidly. If not mentioned potential steps were performed from the open circuit potential to the desired potential with a period of ~1 to 2 min. Subsequently, the potential was returned and maintained at an open circuit to avoid the large accumulation of hydrogen concentration at the interface. In the case of blip charge measurements, the integrated charge under the blip was taken for at least 5-10 blips far from the early double-layer charging of the microelectrode.

In all experiments either deflection or amplitude modulation AFM was used to approach the surface. In the first method, the oscillation amplitude of the cantilever is used as a measure of tip-sample distance. To acquire the cantilever noise power spectral density (PSD) we routed the raw deflection signal from the MFP 3D to a fast digitizer with deep memory. We used a commercial GaGe Octopus-8320 with a 2.5

Ms/s sampling rate, 2,560,000 segment size, and 2,560,000 depth. A 200 Hz high pass filter cut off the DC signal on the raw deflection signal before sending it to the Gage DAQ and a 40 dB gain was applied to maximize the dynamic range of the GaGe DAQ. A 1 MHz low pass filter sets the bandwidth. Cantilever displacement noise power spectral density (PSD) was acquired as follows: at each potential, the raw cantilever deflection signal was measured for 1 second, and a noise PSD was calculated. 30 such noise PSD traces were averaged to represent the averaged deflection noise PSD. While collecting the noise PSD, the piezo drive of the cantilever (amplitude modulation) was turned off and very weak z feedback gains (integral feedback gain 0.01) were used. The average noise PSD was integrated from 220 Hz to 0.6 MHz to get the average RMS deflection noise. The units ($V^2$) are converted to 'pm' using the calibrated optical lever sensitivity (nm/V). The sum signal on the photodiode was continuously monitored to note any changes in the optical path during the hydrogen evolution reaction.

**Estimation of H$_2$ bubble size from chronoamperometry**

Charge under blip is evaluated by integrating the area of transient blip currents for various blips and the resulted in average blip charge ($Q$) was experimentally obtained and provided in Table S1. The double-layer charge density, $q_{DL}$ of the Pt UME was estimated at the potential where there was no hydrogen bubble formation occurring ($E_f$ = 0.3, 0.1, and - 0.1 V). The obtained $q_{DL}$ was linearly extrapolated to the H2 gas bubble formation $E_f$.

$$q_{DL} = Q_{DL}/A_{ecsa} \quad \text{----- (1)}$$

Where $q_{DL}$ is double-layer charge density, $Q_{DL}$ is a double-layer charge, measured from step transient from OCP to $E_F$ as shown in Figure S2a. $A_{ecsa}$ is electrochemically active surface area. $A_{ecsa} = A_{geometrical} \times R$ Where $A_{geometrical}$ is a geometrical area of Pt UME and $R$ is the roughness factor. $R = 2$ is noted for well-polished electrodes.[1]

The two-dimensional projection area of a spherical H$_2$ gas bubble $A_{Proj} = Q/q_{DL}$ ------ (2)

Spherical H$_2$ gas bubble surface area, $A_{surf} = 4A_{proj} = 4\left(QA_{ecsa}/Q_{DL}\right)$ ----------- (3)

Since, $A_{Surf} = 4\pi r^2$, where r = gas bubble radius ------- (4)

From equation (3) and (4), gas bubble radius on the electrode surface = $r = \sqrt{\left(QA_{ecsa}/\pi Q_{DL}\right)}$

**Table S1.** H$_2$ bubble size is estimated from the average blip charge under the current-time transients.

| $E_F$ (V vs. RHE) | Average Blip charge, $Q$ (nC) | Double Layer charge density, $q_{DL}$ (μC/cm²) | Projected area of H$_2$ gas bubble $A_{proj}$ (μm²) | H$_2$ gas bubble surface area unit $A_{surf}$ | H$_2$ gas bubble Radius, $r$ (μm) |
|---|---|---|---|---|---|
| -0.4 | 0.00212 | 760 | 2.79E-08 | 1.12E-07 | 0.942 |
| -0.5 (i) | 0.142 | 860 | 1.65E-07 | 6.60E-07 | 2.29 |
| -0.5 (ii) | 1.2053 | 860 | 1.40E-06 | 5.61E-06 | 6.68 |
| -0.55 | 1.01 | 910 | 1.11E-06 | 4.44E-06 | 5.94 |
| -0.575 | 2.32 | 935 | 2.48E-06 | 9.93E-06 | 8.89 |
| -0.6 | 3.35251 | 960 | 3.49E-06 | 1.40E-05 | 10.5 |

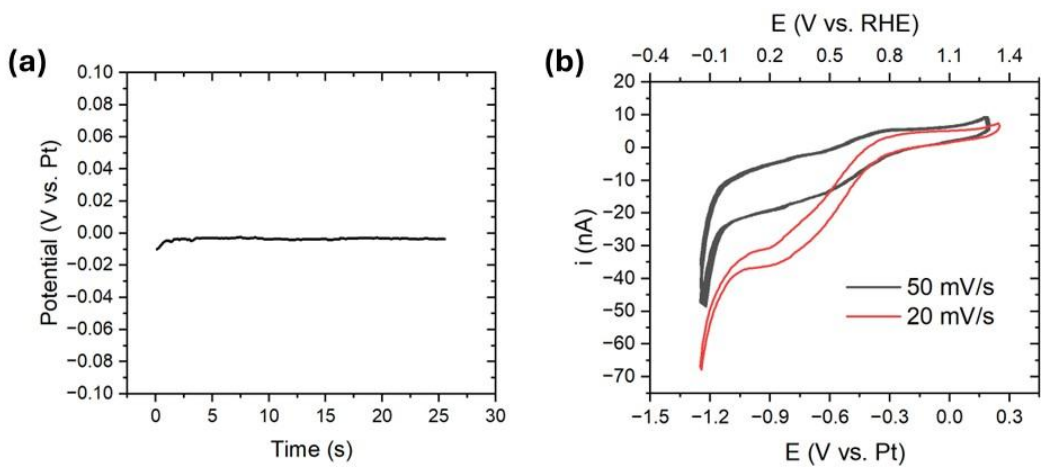

**Figure S1**. (a) Open circuit potential (OCP) of 25 μm Pt UME versus Pt wire (b) CV of Pt UME at different scan rates (50, 20 mV/s) in 0.1 M $K_2SO_4$ in an open atmosphere. The steady-state current observed in the red trace is due to an oxygen reduction reaction.

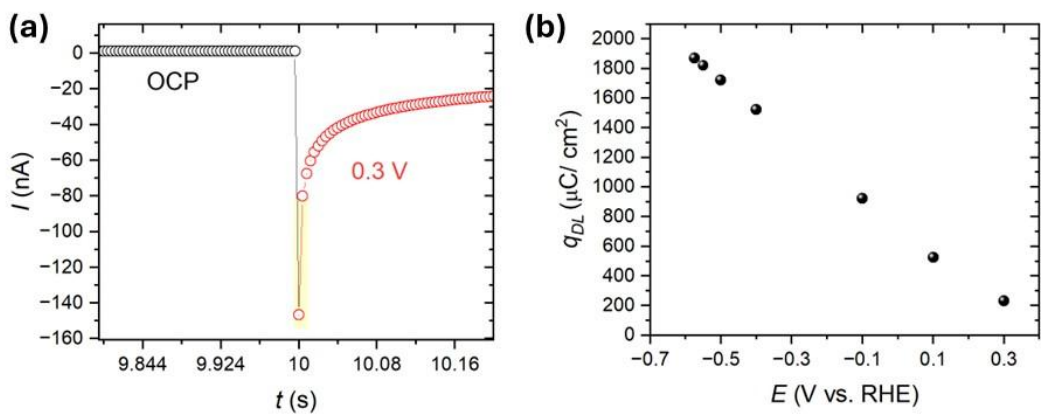

**Figure S2.** (a) A sample plot of double-layer (DL) charge ($q_{DL}$) estimation from the *i-t* curve for the initial 10 ms. (b) Calibration plot for the $q_{DL}$ at different $E_f$ and linearly extrapolated to $H_2$ gas bubble evolution potential. (c) Estimated hydrogen bubble sizes as a function of Pt UME potential.

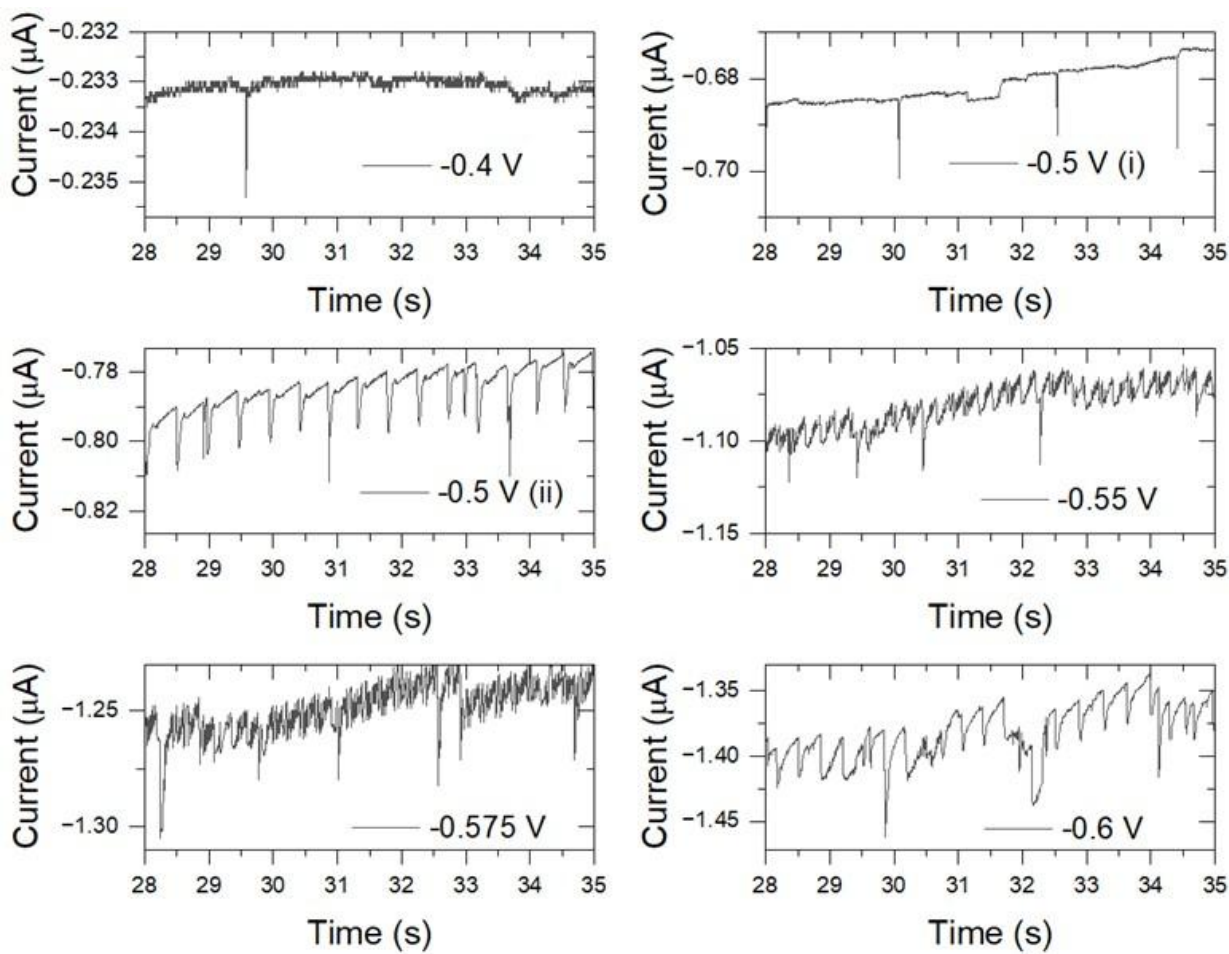

**Figure S3.** Chronoamperograms of Pt UME at $E_f$ = -0.4 V, -0.5 V (i), -0.5 V (ii), -0.55 V, -0.575 V, -0.6 V. -0.5 (i) are first collected and -0.5 V (ii) represents the subsequently collected data at the same potential after -0.5 V.

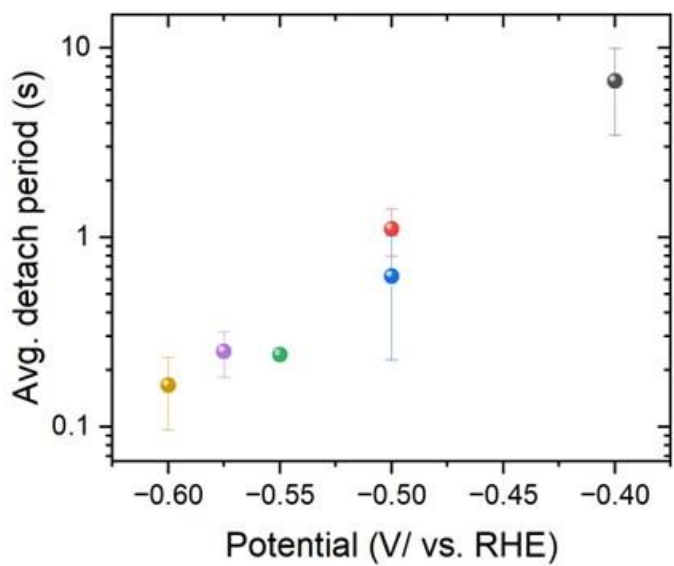

**Figure S4.** (a) Average detachment period of $H_2$ gas bubbles versus $E_f$ of Pt UME.

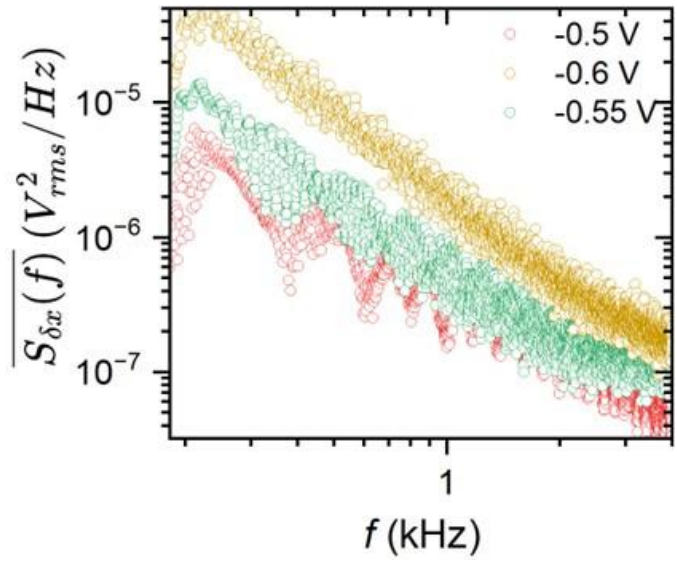

**Figure S5.** Average deflection noise PSD of AFM cantilever recorded during HER at $E_f$ = -0.5, -0.55 V, and -0.6 V potentials in 0.1 M $K_2SO_4$.

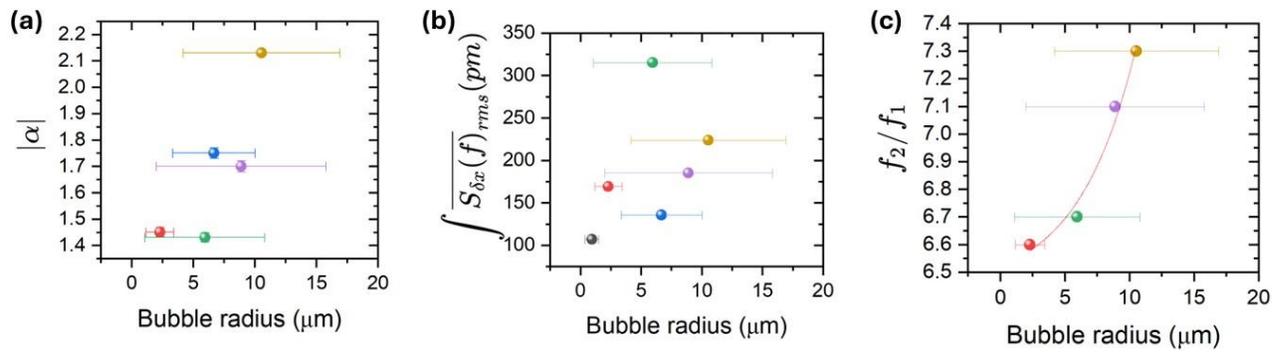

**Figure S5.** Plots of (a) absolute slope ($\alpha$), total force noise ($\int \overline{S_{\delta x}(f)}$), the ratio of resonance frequencies ($f_2/f_1$) of AFM cantilever vs. $H_2$ bubble radius.

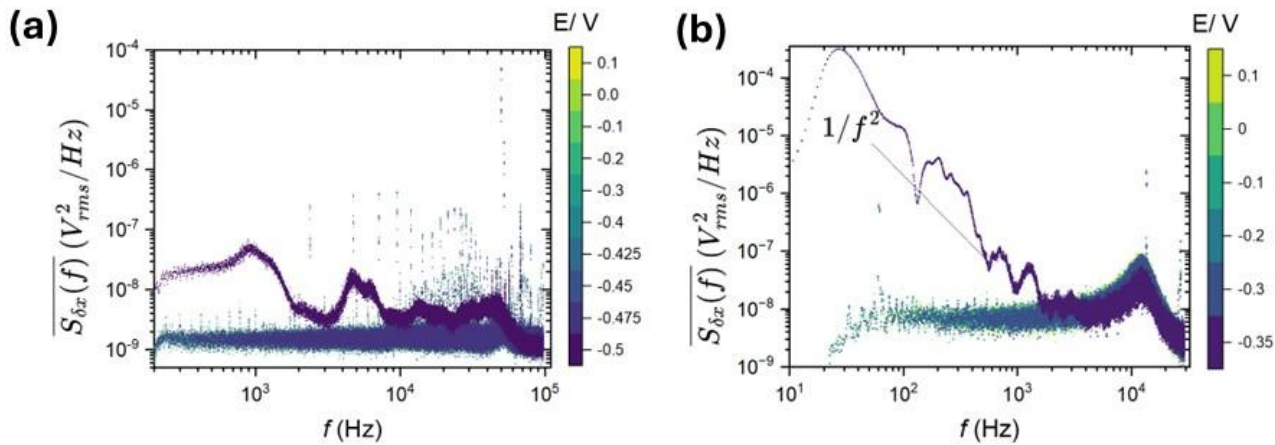

**Figure S6**. Average deflection noise PSD of Si cantilever with (a) spring constant ($k$) = 4.33 N/m and (b) $k$ = 0.3 N/m measured for HER on Pt UME in 0.1 M $K_2SO_4$. Tip-sample distance = 10 nm

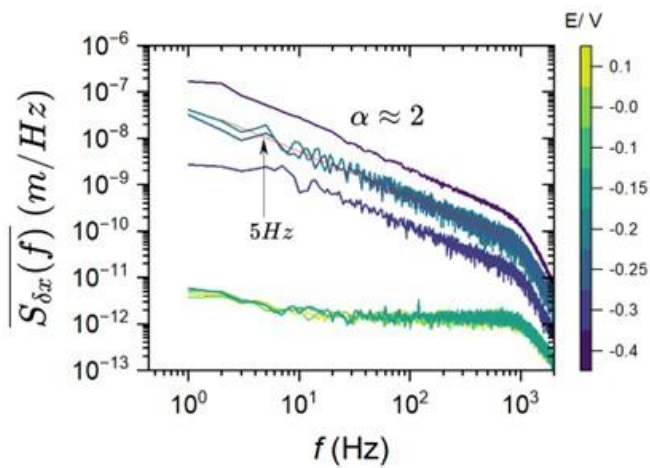

**Figure S7.** Force noise spectral density of deflection signal using tipless Si cantilever ($k$ = 0.22 N/m) at different $E_f$. Oscillation peaks are observed at ∼ 5 Hz.

**Supplementary references**

(1) Bodappa, N. Rapid assessment of platinum disk ultramicroelectrodes' sealing quality by a cyclic voltammetry approach. *Analytical Methods* **2020**, *12* (27), 3545-3550. DOI: 10.1039/d0ay00649a.